\documentclass[useAMS,usenatbib]{mn2e}
\usepackage{graphicx}

\newcommand{\be}{\begin{equation}}
\newcommand{\ee}{\end{equation}}
\newcommand{\bea}{\begin{eqnarray}}
\newcommand{\eea}{\end{eqnarray}}

\title[Empirical Photometric Redshifts of Luminous Red Galaxies and Clusters in SDSS]{Empirical Photometric Redshifts of Luminous Red Galaxies and Clusters in SDSS}
\author[P. A. A. Lopes]{P. A. A. Lopes$^{1,2}$\thanks{E-mail: 
paal@das.inpe.br}\\
$^{1}$Instituto Nacional de Pesquisas Espaciais -- Divis\~ao de
Astrof{\' \i}sica (CEA), Avenida dos Astronautas, 1758\\
S\~ao Jos\'e dos Campos, SP 12227-010, Brasil \\
$^{2}$Harvard-Smithsonian Center for
Astrophysics, 60 Garden Street, Cambridge, MA 02138, USA}
\begin{document}

\date{Accepted  Received ; in original form }

\pagerange{\pageref{firstpage}--\pageref{lastpage}} \pubyear{2007}

\maketitle

\label{firstpage}

\begin{abstract}
In this work I discuss the necessary steps for deriving photometric redshifts
for luminous red galaxies (LRGs) and galaxy clusters through simple empirical 
methods. The data used is from the Sloan Digital Sky Survey (SDSS). I show
that with three bands only ({\it gri}) it is possible to achieve results as
accurate as the ones obtained by other techniques, generally based on
more filters. In particular, the use
of the $(g-i)$ color helps improving the final redshifts (especially
for clusters), as this color
monotonically increases up to $z \sim 0.8$. For the LRGs I generate a catalog
of $\sim 1.5$ million objects at $z < 0.70$. The accuracy of this
catalog is $\sigma = 0.027$ for $z \le 0.55$ and $\sigma = 0.049$ for 
$0.55 < z \le 0.70$. The photometric redshift technique employed for
clusters is independent of a cluster selection algorithm. Thus, it
can be applied to systems selected by any method or wavelength, as
long as the proper optical photometry is available. When comparing the
redshift listed in literature to the photometric estimate, the accuracy
achieved for clusters is $\sigma = 0.024$ for $z \le 0.30$ 
and $\sigma = 0.037$ for $030 < z \le 0.55$. However, when considering
the spectroscopic redshift as the mean value of SDSS galaxies on each
cluster region, the accuracy is at the same level as found by other
authors: $\sigma = 0.011$ for $z \le 0.30$ and $\sigma = 0.016$ for
$030 < z \le 0.55$. The photometric redshift relation derived here is
applied to thousands of cluster candidates selected elsewhere. I have
also used galaxy photometric redshifts
available in SDSS to identify groups in redshift space and then
compare the redshift peak of the nearest group to each cluster
redshift. This procedure provides an alternative approach for cluster
selection, especially at high redshifts, as the cluster red sequence may
be poorly defined.
\end{abstract}


\section{Introduction}

Recent galaxy redshift surveys (2dFGRS, \citealt{col01};
SDSS, \citealt{yor00}) have provided the astronomical community a 
unique view of the local universe ($z \sim 0.1$). Such surveys are based on 
spectrographs that simultaneously observe hundreds of objects. Although the
improvement respective to a decade ago is enormous, larger and mainly deeper
general spectroscopic surveys are not yet possible with current
instrumentation. Note that surveys targeting specific populations,
such as Luminous Red Galaxies (SDSS, \citealt{eis01}) or star-forming 
galaxies (Wiggle-z, \citealt{gla07}), can sample much larger volumes.
For the mean time, photometric redshifts provide a valuable alternative to
probe faint sources within large areas.

Photometric redshift techniques are essentially a mechanism  to convert
photometric properties of galaxies (such as colors) into redshift and 
physical properties (e.g., luminosity and type). Thus, with 
the proper choice of passbands and the use of an accurate 
photometric redshift algorithm, it is possible to map the distant 
universe in three dimensions. These surveys represent a powerful tool
for studying the statistical properties of galaxies and their evolution.

There are several photometric redshift estimators 
developed to date. These can be generally
classified either as empirical or template-based methods. In the first
case a direct relation is obtained through the comparison of the
photometric properties (colors) and spectroscopic redshifts. Such empirical 
relations can be derived, for instance, through polynomial 
fitting \citep{con95} or neural networks \citep{col04}. The template-based
algorithms rely on the availability of a set of galaxy templates. These should
accurately represent the distribution of galaxy SEDs and their evolution with 
look back time \citep{csa03}. Hybrid photometric redshift techniques have
also been proposed in the last few years. They combine 
the advantages of empirical and
template fitting methods by iteratively improving the concordance between
photometric data and the spectral energy distributions. In other words, the
template spectra is reconstructed to best match the observed photometric 
measurements of each galaxy \citep{bud00, csa03}.

Luminous red galaxies are marked by uniform spectral energy distributions,
characterized by a strong break at 4000 {\AA} due to the accumulation of a 
number of metal lines. The shift of this feature through different filters is
strong correlated with redshift. These galaxies are also known to be 
some of the most luminous objects in the universe and are preferentially 
found at high density environments, rendering these objects an 
interesting tool for selecting and studying clusters. All that said, 
it is clear that LRGs comprise an optimal population to derive 
accurate photometric redshifts to very large distances.

This paper describes the construction of a large photometric redshift 
catalog of LRGs at $z < 0.70$. This catalog is based on simple polynomial
fitting of the relations between galaxy colors and spectroscopic redshifts.
I explore the use of different colors from SDSS, showing that with three
bands only it is possible to achieve results comparable to more elaborated
empirical techniques, such as ANNz \citep{col04}, kd-trees or
the nearest-neighbour method \citep{csa03}. In addition, I employ similar
relations to derive photometric redshifts of galaxy clusters.

When estimating redshifts of clusters the main drawback is the need to
apply a background correction when selecting probable cluster galaxies
seen in two dimensions. I discuss different possibilities when minimizing
the background effects, showing that the most precise results can be
achieved when selecting the reddest galaxies (the selection is based
on the $u-r$ color). Photometric redshift estimates of clusters rely
on precise values of their median color. Elliptical or S0 galaxies comprise
the main population in the central regions of galaxy clusters. Thus, 
one would like to use these galaxy types when
estimating the typical colors of clusters. As it is shown in $\S$4.2,
at low redshifts a simple statistical background correction is enough
to minimize the influence of galaxies that do not belong to the
clusters, and to accurately estimate cluster colors. However, at
higher redshifts this simple correction leads to an underestimation of
cluster colors compared to the expected values for ellipticals. That
is also due to the increase with redshift in the fraction of blue
galaxies in clusters \citep{but84}. To circumvent this problem, the
use of the $u-r$ color plays a key role to help the selection of early
type systems and thus reduce the scatter of the observed colors of
clusters. More details are found in $\S$4.2.

This paper is divided as follows. In the next section I describe the SDSS
survey, which is used as the basis for obtaining the empirical relations and 
evaluate the results. In section 3 I describe the selection of LRGs
and the photometric redshift technique employed for these
objects. The same is done for clusters in $\S$4, where I also make
considerations about redshift accuracy. I also use
galaxy photometric redshifts from SDSS for the identification of
groups in redshift space. I summarize the
results in $\S$5. Through this work I assumed a cosmology with
$\Omega_m = 0.3, \Omega_{\Lambda} = 0.7$ and H$_0 = 100$ $h$ $km$
$s^{-1}$ Mpc$^{-1}$, with $h$ set to 0.7.

\section{Data}

The photometric and spectroscopic data for this paper were taken from 
the fifth release of the Sloan Digital Sky Survey \citep{yor00}. 
The SDSS consists of an imaging survey of $\pi$ steradians of the 
northern  sky in five optical passbands ({\it ugriz}), 
from 3,500--8,900 \AA. This will provide photometry for of order $5
\times 10^{7}$ galaxies. Spectroscopic will provide redshifts 
and spectra for  $\sim 10^{6}$ of these. The survey is 
carried  out using a 2.5 m telescope, an imaging  mosaic camera 
with  30 CCDs, two fiber-fed spectrographs and a 0.5 m  telescope 
for the photometric  calibration. The imaging survey is taken in 
drift-scan mode and the data are processed  with a photometric 
pipeline (PHOTO) specially written for the SDSS data.

Targets for spectroscopy are selected by the targeting pipeline from
the imaging. Spectroscopic fibers are assigned to the targets by
a tiling algorithm \citep{bla03}. The minimum distance of 55 arcsec 
between the fibers leads to a loss of $\sim$ 6\% of galaxies, which
is the main source of incompleteness.

The spectroscopic survey is originally divided in three samples. The 'Main', 
flux-limited sample, has a median redshift of 0.104 and a limiting magnitude
of $r_{petro} \sim 17.77$ \citep{str02}. As this limit is much brighter than 
that for the imaging, the redshift completeness is nearly 100\%. The second 
sample is the luminous red galaxy sample (LRG), which is approximately 
volume-limited to $z \approx 0.38$ \citep{eis01}, extending to 
$z \approx 0.55$. Finally, the quasar sample is defined by objects 
with colors distinct from those of ordinary stars. The completeness of 
this sample depends somewhat on redshift. In particular, the completeness 
is low for $2.4 < z < 2.9$, where the quasar and stellar loci cross; it 
is similarly low at redshifts around 3.5 and 4.5. 

In addition to these data, the 2dF-SDSS LRG and 
Quasar Survey (2SLAQ; \citealt{can06}) has been recently completed. 
This survey exploits the high-quality SDSS imaging combined with the 
extraordinary spectroscopic capabilities of the Two-degree Field (2dF) 
instrument on the 3.9-m Anglo-Australian Telescope. It then results in 
a spectroscopic redshift catalog of $\sim$13,000 LRGs at $0.4<z<0.7$. 
These data can be found in the data release five (DR5) of SDSS or
directly from the 2SLAQ website\footnote{http://www.2slaq.info/.}. 
In this paper I use photometric data from SDSS and spectroscopic 
redshifts from SDSS and 2SLAQ surveys (available within SDSS).

All the data selected from SDSS is from the DR5. I have selected only
objects from the ``Galaxy'' view (so that only \texttt{PRIMARY} objects are
allowed) in order to avoid duplicate observations. Standard flags for
clean photometry are also enforced. When selecting spectra and imaging,
a {\it joined} query of the Galaxy and SpecObj (objects with
clean spectra) views is performed. All the magnitudes retrieved
from SDSS are de-reddened (corrected for extinction) model magnitudes.

\section{Selection of Luminous Red Galaxies}

The selection criteria adopted for constructing a photometric sample of LRGs
is analogous to the description given in \citet{pad05}, which is aimed
at selecting a uniform sample of LRGs at $0.2 < z < 0.7$. Two different
criteria are applied for selecting a low redshift sample 
(Cut I, $z < 0.4$) and a high redshift sample (Cut II, $z > 0.4$). Initially, 
two color tracks are defined:

\bea
c_{\perp} \equiv (r-i) - (g-r)/4 - 0.18 \,\,\, ,\\
d_{\perp} \equiv (r-i) - (g-r)/8 \approx r-i \,\,\,.
\label{eq:perpdef}
\eea
Then, the following color cuts are applied
\bea
{\rm Cut\,\,I :} & \mid c_{\perp} \mid < 0.2 \,\,\,;\\
{\rm Cut\,\,II :} & d_{\perp} > 0.55 \,\,\,, \\
& g-r > 1.4 \,\,\,,
\label{eq:colourcuts}
\eea
The final cut, $g-r > 1.4$, is effective on isolating the sample
from the stellar locus. In addition to these selection
criteria, all galaxies with $g-r > 3$ and $r-i > 1.5$ are eliminated.
These last constraints are helpful on removing stars with 
unusual colors, without discarding real galaxies
\citep{pad05}. However, it is important to keep in mind that a 5\%
stellar contamination may still be present, as pointed out by
\citet{col07} (see section 4 of their paper), who applied similar
selection criteria. Nonetheless, their criteria leads to the selection
of more objects than the one adopted here.

These color cuts are still not enough to select LRGs from SDSS (see discussion 
in \citealt{eis01}). Therefore, additional cuts in magnitude are applied. 
First, a color track which is approximately parallel to the low-redshift
locus is defined for Cut I
\be
c_{||} = 0.7 (g-r) + 1.2(r-i-0.18) \,\,\,,
\label{fig:cpllel}
\ee
Then, the following cuts are implemented 
\bea
{\rm Cut\,\,I :} & r_{Petro} < 13.6 + c_{||}/0.3 \,\,\,,\nonumber \\
& r_{Petro} < 19.7 \,\,\,; 
\label{eq:colourmagcuts1}
\eea
\bea
{\rm Cut\,\,II :} & i < 18.3 + 2d_{\perp} \,\,\,, \nonumber \\
& i < 20 \,\,\,.
\label{eq:colourmagcuts}
\eea
$r_{Petro}$ is used for consistency with the original SDSS LRG target selection.
Except for the numerical values of the magnitude cuts in equation 7, Cut I 
is identical to the SDSS LRG Cut I. The numerical values for Cut II are chosen
to derive a population consistent with the first cut. At the redshift range
sampled by Cut II the 4000 {\AA} break is moving through the $r$ band. As a
consequence, the $r$ band $K-$corrections are very sensitive to redshift.
Thus, using the $i$ band for Cut II leads to a more robust selection.

When applying these criteria to select LRGs from the DR5 of SDSS a total 
of 578,160 galaxies are selected using Cut I and 896,988 through Cut II.
The combined sample, after excluding overlapping galaxies, adds to
1,459,536 luminous red galaxies. This sample is from now on called the 
photometric sample. Note that the high redshift sample has
approximately 74\% of the MegaZ-LRG catalog \citep{col07}. This last catalog 
was selected from the data release 4 of SDSS (DR4). If the same criteria 
adopted by \citet{col07} is applied to DR5 the number of LRGs retrieved is 
$\sim$1.4 million. So, the high redshift sample in the current work (Cut II)
actually represents $\sim$ 64\% of the MegaZ-LRG catalog. That is due to the
different criteria employed here. I allow only objects at $1.4 < g-r
\le 3$ and $r-i \le 1.5$, while \citet{col07} uses $0.5 < g-r \le 3$
and $r-i < 2$. Besides that, they select galaxies at $d_{\perp} > 0.5$,
while here the adopted cut is  $d_{\perp} > 0.55$.
When imposing that galaxies should have spectroscopic
measured redshifts and applying the same criteria as above, there are
197,956 luminous red galaxies in SDSS. Out of these, 186,572 are 
at low redshifts (Cut I) and 11,384 at high redshifts (Cut II). The 
small number of galaxies with spectra available at high redshift is 
due to the fact that 2SLAQ was restricted to a small number of fields 
located in the equatorial stripe of the SDSS survey area. This set is called 
the spectroscopic sample. Note that this sample does not
include stars, as those were removed according to their spectroscopic
identification.

\subsection{Photometric redshifts of LRGs}

The empirical photometric redshift estimators rely on the existence of 
a training set of objects with spectroscopic redshifts. This set should
be representative, in terms of photometry and redshift, of the target sample
which will be used later on. The training set used here is the 
spectroscopic sample (with 197,956 objects) mentioned above. Actually, this
sample is divided into a ``training'' and ``evaluation'' samples.
I randomly selected 10,000 objects out of the 197,956 LRGs
to be the ``training'' set. The remaining 187,956 galaxies are kept to
play the role of an ``evaluation'' sample. The training set is then
used to derive an empirical relation between galaxy colors and redshift,
which is then applied to the the evaluation set. I found that 
increasing the training sample to 20,000 objects does not represent a 
meaningful gain in accuracy. Forcing the ``random'' selection to have
a fixed percentage at $z > 0.40$ (say 60$\%$) also leads to similar results.

On what follows I discuss which colors are best suited for deriving an
empirical relation used for photometric redshift estimates. In Figure 1
I show the variation with redshift of the apparent magnitude $r$ and
of five SDSS colors, namely ({\it u-g}), ({\it g-r}), ({\it g-i}),
({\it r-i}), ({\it i-z}). The data points represent the 10,000
galaxies randomly selected for the training sample. 
On each bandpass K-corrections are obtained through the 
convolution of a SED characteristic of early type galaxies 
(taken from \citealt{col80}; CWW from 
now on) with the SDSS filters. The expected colors 
in different redshifts are the result of adding the color of an elliptical
galaxy at zero redshift and the difference in K-corrections between two
bands. The zero redshift colors are taken from table 3 of \citep{fuk95}.
These color tracks are shown by the solid lines (red in the electronic edition)
of panels (b)-(f) of Figure 1. A small offset was noticeable between the color
tracks and the data points. I estimated these offsets (a factor $< 0.15$) and
took them in account for the figure. In panel (d) the dotted line (green in
the electronic 
edition) indicates a second order polynomial fit to the relation 
between ({\it g-i}) color and redshift. In this panel the dashed line 
(blue in the electronic figure) shows the result of a fourth 
degree polynomial.

\begin{figure*}
\begin{center}
\leavevmode
\includegraphics[width=6.0in]{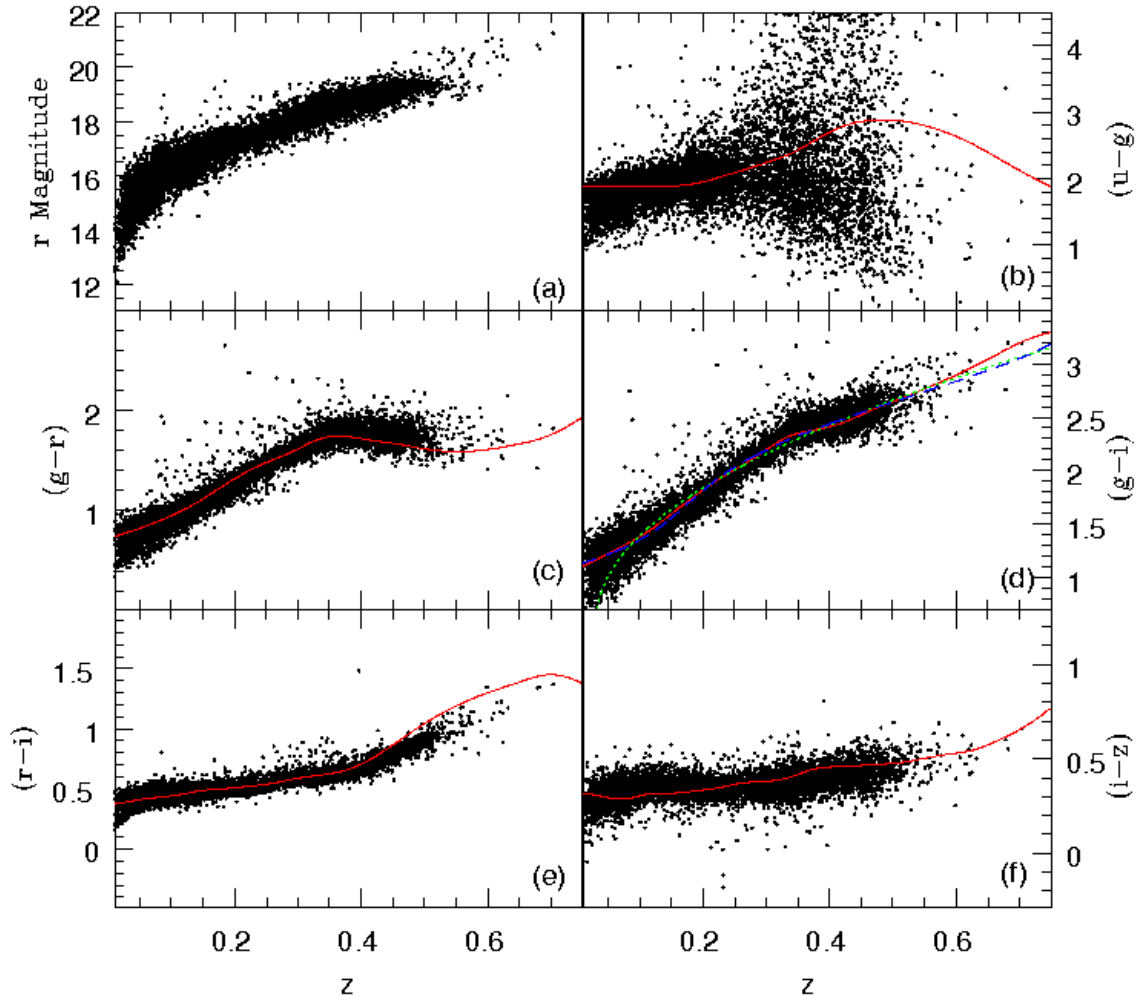}
\end{center}
\caption{The variation of apparent magnitude $r$ and five SDSS colors with
redshift. The six panels show the following parameters versus redshift: (a) 
magnitude; (b) ({\it u-g}); (c) ({\it g-r}); (d) ({\it g-i}); 
(e) ({\it r-i}); (f) ({\it i-z}). The solid line (red  in the electronic 
version) on each panel indicates the expected color variation of 
early type galaxies. In panel (d) the dotted line (green in the electronic 
edition) indicates a second order polynomial fit to the relation 
between ({\it g-i}) color and redshift. In the same panel the dashed line 
(blue in the electronic figure) shows the result of a fourth 
degree polynomial.}
\label{fig:colorsredshift}
\end{figure*}

A few features are readily noticed from the inspection of this figure. First,
({\it u-g}) shows a large scatter and does not follow the color track 
expected for an elliptical galaxy (due to the lower sensitivity of the
$u$ filter, specially for LRGs). Second, the ({\it i-z}) color shows
little variation with redshift. Thus, we do not expect these two colors 
to contribute in a meaningful way for a photometric redshift estimator.
Third, the ({\it g-r}) shows the expected large variation at low redshifts
($z < 0.35$), becoming nearly flat afterwards. 
The ({\it r-i}) color shows the opposite
trend, being nearly constant at $z < 0.35$ and increasing fast for higher-z.
The ({\it g-i}) combines the results of the two previous colors, showing
a large variation at low-z and a less steep dependence at higher-z. These 
results are mainly associated to the shifting of the 4000 {\AA} break between
the $g$ and $r$ filters at $z \sim 0.35$. Last, it is worth to note
that the ({\it r-i}) color has a much smaller scatter in comparison to
the ones based on the $g-$band ({\it u-g, g-r, g-i}). That happens
cause the reddest filters are better suited for sampling these types
of galaxies.

I then used the training data to estimate different empirical relations 
between colors and redshift. These relations are based on a variety of 
combinations of colors, with the use of the $r$ band magnitude in a few cases.
These relations are then applied to the 187,956 galaxies of the evaluation
sample. The redshift accuracy is characterized by the residual between
the spectroscopic and photometric redshifts ($\delta_z = z_{spec} -
z_{phot}$) and the standard deviation
\bea
{\rm \sigma} = {\sqrt{\frac{1}{N-1} \sum({\delta_{zi} - \mu)^2}}} \,\,\,\,\,,
\label{eq:standarddeviation}
\eea
where $\delta_{zi}$ is the residual for the i-th galaxy and
$\mu = <\delta_z>$ is the mean residual. The sum is performed over all N data
points. I also computed the mean and standard
deviation (now called $\mu_0$ and $\sigma_0$) for the case where the residual 
is weighted by the factor $1+z_{spec}$ 
($\delta_{z0} = (z_{spec} - z_{phot}$)/($1+z_{spec}$)). When computing
$\mu$ and $\sigma$ (or $\mu_0$ and $\sigma_0$) I only use galaxies with
$|\delta_z|$ or $|\delta_{z0}| < 0.10$. This gives equivalent results to
reject outliers at the 3-$\sigma$ level. Including the gross outliers
increases the standard deviation, as $\sigma$ is very sensitive to the
presence of outliers. For instance, in the fifth row of Table 1
(results based in the colors g-r, g-i and r-i) the values of $\sigma$
and $\sigma_0$ are raised from 0.027 and 0.023 to 0.032 and 0.026,
respectively.

The mean and standard deviation, as
well the fraction of galaxies with a valid $z_{phot}$ (the number of
gross outliers is 100 minus this fraction), 
obtained for different empirical relations are summarized in Table 1. The
last row of this table has the results obtained when considering the color 
track based on the elliptical template from \citet{col80} for 
the ($g-i$) color. In other words, I simply use the track exhibited as a
solid line in panel (d) of Figure 1 to compute redshifts from the observed
(g-i) color. The remaining rows (1-7) in Table 1 show the results for 
the following polynomial fits:
\bea
\lefteqn{ z_{phot} =  {A + B(g-i) + C(g-i)^2 + D(g-i)^3 + E(g-i)^4}, } \, \nonumber \\
\lefteqn{ z_{phot} =  {A + B(g-i) + C(r-i)}, } \, \nonumber \\
\lefteqn{ z_{phot} =  {A + Br + C(g-r) + D(r-i)}, } \, \nonumber \\
\lefteqn{ z_{phot} =  {A + Br + C(g-i) + D(r-i)}, } \, \nonumber \\
\lefteqn{ z_{phot} =  {A + B(g-r) + C(g-i) + D(r-i)}, } \, \nonumber \\
\lefteqn{ z_{phot} =  {A + B(g-r) + C(g-i) + D(r-i) + E(r-z)}, } \, \nonumber \\ 
\lefteqn{ z_{phot} =  {A + B(g-r) + C(g-i) + D(r-i) + E(i-z)} } \, \nonumber \\
& & {}+ F(g-r)^2 + G(g-i)^2 + H(r-i)^2 + I(i-z)^2 \,. \nonumber
\label{eq:colourmagcuts}
\eea

\begin{table}
\begin{tabular}{lccccc} 
\hline
\multicolumn{6}{c}
{Photometric redshift errors of LRGs from different relations} \\ 
\multicolumn{6}{c}
{} \\
Relation & $\mu$ & $\sigma$ & $\mu_0$ & $\sigma_0$ & Fraction(\%) \\ 
\hline
gi4 & -0.0024 & 0.029 & -0.0026 & 0.025 & 94.1 \\
giri & -0.0008 & 0.028 & -0.0012 & 0.024 & 97.1 \\
rgrri & -0.0007 & 0.027 & -0.0008 & 0.023 & 96.4 \\
rgiri & -0.0007 & 0.027 & -0.0008 & 0.023 & 96.5 \\
grgiri & -0.0075 & 0.027 & -0.0068 & 0.023 & 97.4 \\
grgirirz & -0.0068 & 0.027 & -0.0062 & 0.024 & 97.1 \\
grgiriiz2 & -0.0062 & 0.028 & -0.0058 & 0.025 & 96.9 \\
TEMP-CWW & 0.0016 & 0.030 & 0.0013 & 0.026 & 99.1 \\

\hline
\end{tabular}
\caption{The mean, standard deviation and fraction of LRGs with a valid
photometric redshift ($z_{phot} > 0$ and $|\delta_{z0}| < 0.10$). All rows 
show the results of polynomial fits obtained using different parameters. 
The first row lists the results when applying a polynomial of fourth 
order to the relation between the color ($g-i$) and redshift. In the second 
row I list the results when using the colors ($g-i$) and ($r-i$). Those based 
on the $r$ magnitude and colors ($g-r$) and ($r-i$) are show in the third row. 
In the fourth row the results represent the use of the $r$ magnitude 
and colors ($g-i$) and ($r-i$). The fifth row has the results when using
the colors ($g-r$), ($g-i$) and ($r-i$). In the sixth row I show the results
obtained when adding the color ($r-z$) to the previous set. The seventh has
the result of a second order polynomial to the colors ($g-r$), ($g-i$), 
($r-i$) and ($i-z$). Details about the fits are given in the text. In the
last row I show the results achieved when using the color track based on
the elliptical template from \citet{col80} for the ($g-i$) color.}
\label{tab:errors_lrgs}
\end{table}

From the inspection of Table 1 we see no large differences among the different
relations. It is interesting to note that the results based on the ($g-i$)
color track (using the SED from CWW) show the highest fraction of galaxies with
a valid photometric redshift. However, the scatter determined in this way is 
a little larger when compared to other solutions. Besides that fact, the 
inspection of the relation $z_{phot} \times z_{spec}$ shows a poor correlation 
for $z_{spec} > 0.30$ (for the TEMP-CWW results). We also note that
the results based on the $r$ band magnitude and colors have more
outliers than the solutions based on two or more colors (but no
magnitude). The result
based on the $r$ magnitude and colors ($g-r$) and ($r-i$) is at the
same level of the one that uses ($g-i$), instead of ($g-r$). So, for
LRGs, the ($g-i$) does not seem to be superior to the ($g-r$) for
redshift estimation. That is not true for galaxy clusters though ($\S
4$). I have also tried second and fourth order polynomial fits to the
($g-i$) color only. However, the overall results were not better. In
Table 1 we can see that the fraction of outliers increases for this
($g-i$) fourth degree solution. Finally, it is interesting to see that
the use of the $z$ filter
trough the $r-z$ or $i-z$ colors does not help improving the
results. The scatter is at the same level when adding the $r-z$ color,
increasing a little if the $i-z$ is employed with a second order fit.

I also estimated the uncertainty in the photometric redshift estimates,
of each galaxy, through propagation of errors. In this process I 
consider the error in the
coefficients of the empirical fits, as well as the photometric errors
for each magnitude or color. These uncertainties are estimated as shown in the
equation below, where I use the fractional uncertainties in the coefficients
(such as A, B, ...) and in the color and magnitudes used on each fit ($r, 
g-r, g-i$, ...)
\bea
{\rm \Delta z_{phot}} = z_{phot}{\sqrt{\left( \frac{\Delta_A}{A} \right)^2 + \cdots + \left( \frac{\Delta (g-r)}{(g-r)} \right)^2 + \cdots }}\,\,.
\label{eq:errorequation}
\eea

After inspecting Table 1 and plots of the $z_{phot} \times z_{spec}$ relation
I decided to consider as the final catalog of photometric redshifts the
one based on the ($g-r$), ($g-i$) and ($r-i$) colors. The coefficients
for the empirical relation derived for this case are
\bea
\lefteqn{ A = -0.3068 \pm 0.0006} \, \nonumber\\
\lefteqn{ B = \,\,\,\, 6.2005 \pm 0.1333} \, \nonumber\\
\lefteqn{ C = -5.9933 \pm 0.1331} \, \nonumber\\
\lefteqn{ D = \,\,\,\, 6.4932 \pm 0.1324} \, \nonumber
\label{eq:coefficients}
\eea
This decision was mainly motivated by the fact that this relation is the one
to produce fewer outliers and to show small individual galaxy redshift errors. 
The results based on relations that involve the magnitude $r$ have large
galaxy error estimates, due to the large 
fractional uncertainties in the magnitude
coefficient ($\frac{\Delta_B}{B}$). That is the main reason for not
adopting one of these relations. A plot with the comparison between 
$z_{phot}$ and $z_{spec}$ obtained with the ($g-r$), ($g-i$) and ($r-i$)
colors is shown on Figure 2. Here I plot 15,000 randomly selected points from
the evaluation sample (comprising 187,956 galaxies). It is important 
to mention that these results are in good agreement to what has 
been found by other authors \citep{pad05, col07} and the residuals 
show no systematic trends with $z_{spec}$. When considering only galaxies 
with $z_{spec} \le 0.55$, $\sigma = 0.027$ and $\sigma_0 = 0.023$, while
for galaxies at $0.55 < z_{spec} \le 0.70$, $\sigma = 0.049$ 
and $\sigma_0 = 0.040$.

\begin{figure}
\includegraphics[height=0.50\textwidth,angle=0]{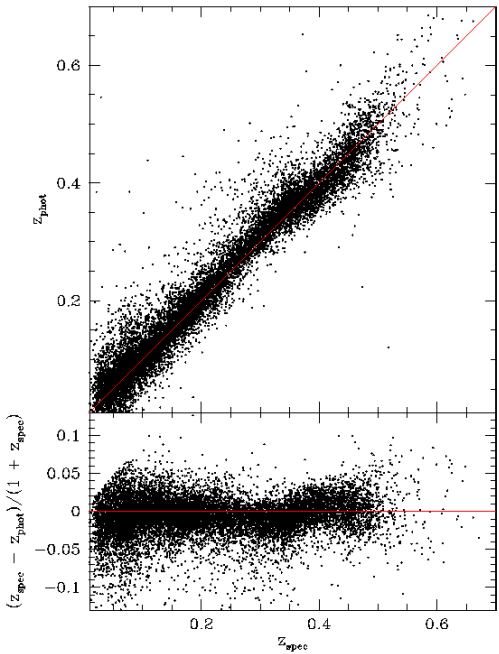}
\caption{Comparison between $z_{phot}$ and $z_{spec}$ using the empirical
relation based on the ($g-r$), ($g-i$) and ($r-i$) colors (upper panel). 
In the lower panel the residuals 
$\delta_{z0} = (z_{spec} - z_{phot}$)/($1+z_{spec}$) are shown. The solid
lines (red in the electronic edition) indicate the Y = X result on 
top and the zero residual in the bottom panel.}
\label{fig:zspeczphot}
\end{figure}

Table 2 presents the luminous red galaxy catalog derived from the
relation based on ($g-r$), ($g-i$) and ($r-i$) colors. The parameters 
listed are \texttt{ra, dec, u, g, r, i, z, u\_err, g\_err, r\_err,
  i\_err, z\_err, z\_phot, err\_z\_phot} and \texttt{objID} (the
object ID within SDSS). The 
magnitudes are the de-reddened model magnitudes. When the estimator
led to a negative photometric redshift I set \texttt{z\_phot and 
err\_z\_phot} to -9.99. That happens for 8,509 galaxies (only  
$0.6 \%$ of the total catalog).

\begin{table*}
\begin{tabular}{ccccccccccccccc} 
\hline
\multicolumn{15}{c}
{Example of the luminous red galaxy catalog} \\ 
\multicolumn{15}{c}
{} \\
\scriptsize ra & \scriptsize dec & \scriptsize u & \scriptsize g & \scriptsize r & \scriptsize i & \scriptsize z & \scriptsize u-err & \scriptsize g-err & \scriptsize r-err & \scriptsize i-err & \scriptsize z-err & \scriptsize z\_phot & \scriptsize err-zp & \scriptsize objID \\
\hline
\scriptsize 0.001571 & \scriptsize 14.982669 & \scriptsize 26.399 & \scriptsize 22.074 & \scriptsize 20.319 & \scriptsize 19.488 & \scriptsize 19.079 & \scriptsize 0.667 & \scriptsize 0.159 & \scriptsize 0.050 & \scriptsize 0.039 &  \scriptsize 0.126 & \scriptsize 0.47247 & \scriptsize 0.06714 & \scriptsize 587730774425600743 \\
\scriptsize 0.002045 & \scriptsize -9.793661 & \scriptsize 22.227 & \scriptsize 21.282 & \scriptsize 19.469 & \scriptsize 18.829 & \scriptsize 18.490 & \scriptsize 0.609 & \scriptsize 0.083 & \scriptsize 0.025  & \scriptsize 0.023 & \scriptsize 0.073 & \scriptsize 0.38900 & \scriptsize 0.03420 & \scriptsize 587727179523227851 \\
\hline
\end{tabular}
\caption{Example table of the LRG catalog containing 1,459,536 objects. 
The magnitudes listed are the de-reddened model magnitudes. 
The full table is available on-line or upon request to the author.}
\label{tab:errors_lrgs}
\end{table*}

\section{Galaxy Clusters}

The knowledge of clusters redshifts is essential for estimating other
physical parameters of these systems (such as luminosity and
richness). Accurate redshifts are also crucial for large scale
structure studies. Perhaps the oldest and simplest way to get
photometric redshifts is through the use of single band galaxy magnitudes
within the cluster region. However, the use of colors turned the photometric
redshifts of clusters much more accurate in the last years. 
Color based techniques
explore the fact that clusters have a large population of early type
galaxies which are characterized by a strong break at 4000 {\AA}. The
observation of this feature through different filter shows a well defined
correlation with redshift.

Some cluster selection methods explore this characteristic, estimating
a cluster redshift at the time of detection \citep{gla05, koe07},
while others do the estimation independently of the selection
procedure \citep{gal03}. Typical accuracy of these color based methods
is $< 10\%$. 

The catalogs from \citet{got02} and \citet{koe07} are based on SDSS. 
So, their results provide an optimal basis for comparison to what I find
here. Two of the goals aimed in this work are: the derivation of
redshifts independent of the selection procedure; and the extension of
these estimates to $z \sim 0.55$, when using SDSS data to $r = 21$. 
The first point is motivated by the fact that some methods for
cluster selection (such as the matched filter or the maxBCG
technique) provide a redshift estimate as the output of the selection
procedure. Although that is a great advantage, it also makes it
difficult to use these methods for redshift estimation of clusters
that do not properly {\it fit} the required properties imposed by these
techniques. The photo-z technique presented here is applicable to
clusters detected by different methods.
I also noticed that none of the SDSS cluster catalogs available in the
literature have redshift estimates to $z \sim 0.55$. The catalog of
\citet{koe07} focus the regime at $0.1 < z < 0.3$
(some other catalogs in SDSS also do not go deeper than
that). However, the work of \citet{kim02} and \citet{got02} use
galaxies at $r \le 21$ and $r \le 21.5$, respectively; and these
catalogs are not intentionally driven to a low redshift regime. The
redshift estimates from \citet{kim02}, based on a matched filter (MF)
technique were intentionally truncated at $z = 0.5$ (as they ran the MF
up to this redshift). The results from
\citet{got02} use the $g-r$ color track for redshift estimation, but
as seen from their figure 14, their photo-zs are truncated at $z \sim
0.44$, while the spectroscopic sample used for comparison goes to $z =
0.5$ and they probably detect higher redshift systems (considering the
magnitude limit adopted). In other
words, although accurate they underestimate the redshifts (at least in
the high-z regime). 

\subsection{Selection of a Training Sample of Galaxy Clusters}

For the determination of photometric redshifts of galaxy clusters the
first step that should be taken is the compilation of a list of
objects with measured spectroscopic redshifts. Unfortunately, there
are not so many clusters with spectra taken at $z > 0.3$, which biases
our sample to low-$z$ clusters. Our calibration sample consists of 512
clusters over the area covered by SDSS (DR5). These come from
\citet{str99, hol99, vil98, cal96} and \citet{mul03}. The combined sample of
these references contains 1805 clusters in the whole sky. After selecting all
clusters with redshifts at $0.02 \le z \le 0.55$ and outside $70.0 <
\alpha < 110.0$ or $270.0 < \alpha < 300.0$ we are left with a list of
1055 clusters. The right ascension limits are meant to avoid most
systems outside the SDSS region. However, many systems that do not
overlap with SDSS are still allowed in this list. Then I
select data from SDSS for all these clusters. Those falling off
the SDSS limits will obviously contain no galaxies. For the remaining
I generate finding charts, which are inspected to see if the regions
around each cluster (8.0 Mpc x 8.0 Mpc) are well sampled (I exclude
clusters with excised regions near their centers). The final list
comprises 512 systems. The sample size may not be large enough to properly
include evolutionary effects. However, the dominant population of red
galaxies in clusters is believed to evolve passively with redshift. So, for
clusters, the average color estimated from these galaxies should provide 
a clear correlation with redshift. In the future it will be interesting to
use larger training samples of clusters for empirical photometric redshift
estimators.

The next step is to investigate which colors provide the best
connection to spectroscopic redshifts. When doing that I noticed that
the relations between magnitude or colors to redshift are very well
established. However, caution should be taken for the background
correction ($\S$4.2) and also to the selection of clusters
which will be used as the evaluation sample. I noticed that most of
the 512 clusters can be well represented by the color tracks of
elliptical galaxies. However, there are a few outliers, specially at
low redshifts, which should definitely be avoided when training a
photo-z estimator. These are mainly associated to wrong redshifts
(clusters with few galaxies with available spectra) or projection
effects (if there is one or more clusters aligned to a low redshift
system the color inferred will probably be wrong). 

I then decided to gather further information from NED to exclude
clusters with a small number of galaxies with redshift available.
When doing that I select from NED all galaxies within 3
arcmins of each cluster center. After inspecting the information
retrieved for each cluster I kept only those with at least 3 galaxies
with a concordant redshift. Besides that, NED also provides a ``special
note'' for some clusters, meaning that there is some peculiarity with
the object (most of times it is a double system or there are different
redshifts listed for it). Most of the 512 clusters are from
\citet{str99} who also gives the number of galaxies used for measuring
the redshift (may be different from above, as the aperture is not
3$'$). I then impose that the clusters should not have the ``special
note'' in NED and have at least 3 galaxies in \citet{str99}. The final
training set comprises 132 systems. Nearly all high redshift clusters
are kept ($z > 0.4$).

\subsection{Magnitudes, colors and background correction}

In this section I show how I compute mean magnitudes and median colors
and apply a background correction along this process. Figure 3 shows 
the color tracks for elliptical (E) galaxies (upper panel)
and late type (Sbc) galaxies (bottom panel) obtained from the
convolution of the CWW
templates with the SDSS filters. The colors shown are ($u-g$),
($u-r$), ($g-r$), ($r-i$), ($i-z$). Other colors are not exhibited to avoid
confusion. The analysis shown here is complimentary to the discussion
done for the LRGs ($\S 3.1$). From the top panel it is easy to
conclude that the
($i-z$) color provides no meaningful information for redshift
estimation, while ($r-i$) is expected to be a powerful redshift
discriminator at $0.40 < z < 0.60$ and ($g-r$) works fine at $z <
0.40$. The two colors based on the $u$ magnitude ($u-g$ and $u-r$)
show a steep variation with redshift at $z < 0.40$ (with $u-g$ being
flat at $z < 0.20$). However, these two colors have a large scatter
($\S 3.1$) due to the lower signal to noise of the $u$ band, specially
for early-type galaxies, rendering their use for redshift estimation
not relevant. But that does not prevent us from using the $u-r$ to
separate early and late type galaxies (see below). 

All mentioned so far is in line with the discussion in section 3.1
(about Figure 1). Perhaps the most important information that could be
extracted from Figure 3 is a way to discriminate early and late type
galaxies. We see that the colors ($g-r$), ($r-i$), ($i-z$) are not good
for that purpose as the tracks of Es and Sbcs are too close and
even overlap in some redshift ranges. The other two colors ($u-g$ and
$u-r$) seem to be useful for this separation. However, we can note
that at low redshifts ($z \sim 0.15$) the difference between Es and
Sbcs is approximately 0.50 mag only for the $u-g$ color, while
something similar to that does not happen for the $u-r$ color at a
fixed redshift. But it
is important to mention that the difference between the lowest value
of the $u-r$ color for ellipticals (at $z = 0$) and the highest for
late types (at $z \sim 0.32$) is of 0.40 mag. So, when using the $u-r$
color a high redshift blue galaxy could be mistaken for a low redshift
red galaxy. Even considering that and the already mentioned large
scatter of these colors I find that they are helpful
for a rough discrimination between galaxy types, which is enough for
the purpose of this work. I adopted ($u-r$) as it shows better
results and for being well know for having a bimodal color
distribution \citep{str01, dri06}.

\begin{figure}
\includegraphics[height=0.50\textwidth,angle=0]{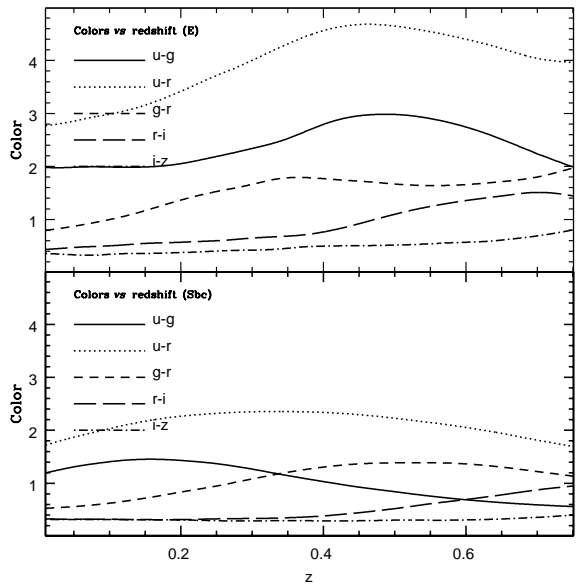}
\caption{Color variations as a function of redshift. The top panels show
color tracks for an elliptical SED, while the bottom panel is for a late type
SED (Sbc). Both SEDs are from CWW. The colors exhibited are: ($u-g$)
with solid line; ($u-r$) as dotted line; ($g-r$) as short dashed;
($r-i$) with the long dashed line and ($i-z$) as the dot-dashed line}
\label{fig:zspeczphot}
\end{figure}

After this discussion we are now ready to compute mean magnitudes
($r-$band) and median colors
($g-i$ and $r-i$, for instance) for all the 132 clusters of the
training sample. For each cluster the number of galaxies as a function of
magnitude (N$_r$) and colors (N$_{(g-i),(r-i)}$) is determined (within
and aperture of 0.50 $h^{-1}$ Mpc, or equivalently 0.71 Mpc for h = 0.7). I use
bins of 0.10 mag for generating histograms of these counts. Blank
fields are used to estimate the background counts. Fifty random
regions (with $0.5^\circ$ radius) are selected in the sky. When
computing the background counts I exclude boxes with values
outside the boundary determined by the mean $\pm$ $3\sigma$ of the 50
original boxes. For each magnitude or color bin the final background
value is the mean of counts from the valid background boxes. Then I
generate the background magnitude and color 
distributions in the same way as done for each cluster, but scaling the
counts for the cluster area. All clusters have their magnitude and
color distributions corrected from the background histograms, leading to
the net cluster histograms. These are used to compute the mean
magnitude and median colors of the clusters. 

I tested three different possibilities for obtaining these corrected cluster
counts. The procedure is executed exactly as described above, but in
the first case I consider all galaxies with the $r$ band magnitude less than
21 ($m_r \le 21$, which is approximately the star/galaxy separation
limit of SDSS) and
in the second I count only galaxies with $m_r \le 21$, but with $m^*-3
\le m_r \le m^*+3$, where $m^*$ is the apparent characteristic
magnitude of clusters. I consider the bright end values of the
double-Schechter cluster luminosity function (LF) obtained by
\citet{pop06}. They found $\alpha = -1.09$ and $M^* = -20.94$ within
$R_{200}$. This value of M* is converted to the same cosmology
used here and to the proper value at $z = 0$ (taking the mean
redshift of their sample as $z = 0.1$). Then, for different redshifts I
adopted an evolutionary correction to the value of M*, given by
$M^*(z) = M^*(0) + Qz$, with Q = -1.4 \citep{yee99}.
The absolute characteristic magnitude ($M^*$) is converted to $m^*$
through the application of the distance modulus formula to each cluster
redshift (the variation of $m^*$ with redshift is shown in the
upper panels of Figure 4). The third test done is to compare the
values obtained
when considering all galaxies to the results found when imposing a selection
according to the $(u-r)$ color.

In Figure 4 I show the dependence of mean apparent magnitude
$r_{mean}$ (two upper panels) and two SDSS colors (median values; four
lower panels) with redshift. The colors exhibited are {\it g-i} and
{\it r-i}. In the top panels the solid line represents the expected
variation of the apparent characteristic magnitude of clusters, while
in the four lower panels the solid line indicates the expected color
variation of E galaxies. In all panels, every black dot is a
cluster (the 132 of the training sample), while the triangles with
error bars (red in the electronic
edition) indicate the mean values in redshift bins of 0.05. The left
panels (a, b and c) show the results for the case where counts are
taken at $m^*-3 \le m_r \le m^*+3$ and a color cut (in {\it u-r}) is
applied for selecting galaxies. The exact values for this cut are
chosen after comparing the {\it u-r} color tracks of Es and Sbcs in
figure 3. Initially, I selected only galaxies with $u-r \ge
2.40$, which already led to very good results. However, I found that these
could still be improved if a variable cut (with redshift) was
applied. The choice of this cut affects most the higher redshift
clusters ($z > 0.20$). I finally decided to select only the galaxies with
$u-r \ge 2.00$ for clusters at $z \le 0.20$, $u-r \ge 2.30$ at $0.20 <
z \le 0.40$ and $u-r \ge 2.45$ at $z > 0.40$. I found these cuts to
give the most accurate photometric redshifts. For comparison
\citet{str01} find that early and late types can be well separated by
a simple color cut at $u-r = 2.22$. 

In the right panels of figure 4 I
show the results of not applying one of the constraints mentioned
above (the fixed luminosity range $m^*-3 \le m_r \le m^*+3$; or
the color cut in {\it u-r}). Panel (d) shows the results for
$r_{mean}$ when using all galaxies at $r \le 21$ (counts are not
restricted to a fixed luminosity range, but the {\it u-r} color cut is
still enforced). Finally, panels (e) and (f) have the median color
variations when I do not impose a color cut in {\it u-r} (but the
fixed luminosity range is still applied).  

\begin{figure*}
\begin{center}
\leavevmode
\includegraphics[width=5.0in]{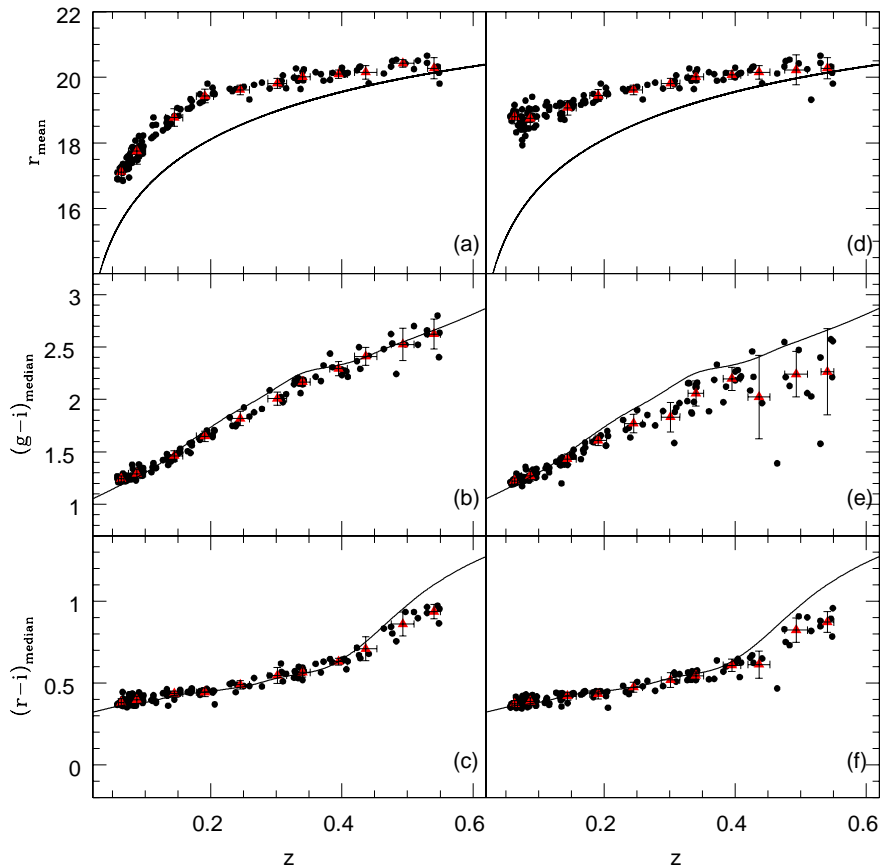}
\end{center}
\caption{The variation of mean apparent magnitude $r_{mean}$ and two
  SDSS colors (median values) with redshift. 
The top two panels show the $r_{mean}$
  variation, while the lower four panels exhibit the median color
  variations ({\it g-i} in the two middle panels and {\it r-i} in the
  lower two). The left panels (a, b and c) have the results when
  considering counts at $m^*-3 \le m_r \le m^*+3$ and with a color cut
  ({\it u-r}) applied for galaxy selection (this cut is explained in
  the text). Panel (d) shows the results for the mean magnitude when
  using all galaxies at $r \le 21$ (counts are not restricted to a
  fixed luminosity range, but the {\it u-r} color cut is still
  applied). Finally, panels (e) and (f) have the median color
  variations when I do not impose a color cut in {\it u-r} (but the
  fixed luminosity range, $m^*-3 \le m_r \le m^*+3$, is still
  enforced). The solid line on each panel indicates the expected color
  variation of early type galaxies (except for panels (a) and (d)
  where they indicate the expected variation of the apparent
  characteristic magnitude of clusters). Each black dot represents one
  of the 132 clusters of the training sample, while the triangles (red
  in the electronic edition) with error bars indicate the mean values
  in redshift bins of 0.05.}
\label{fig:colorsredshift}
\end{figure*}

It is worth mentioning that the further constraints applied above (in
luminosity and color) are intended to improve the background
correction, which is done by the subtraction of every cluster
histogram (in magnitude and colors) by the background
distribution. From panel (a) we can see that the mean cluster
magnitudes show a strong variation up to redshift $\sim 0.4$. After
that the relation tends to become flatter. That is due to the magnitude limit 
considered for the survey ($r = 21.0$), which renders the cluster counts at 
$m^*-3 \le m_r \le m^*+3$ truncated for high redshift systems (as
$m^*+3$ extrapolates $r = 21.0$). Applying a correction to the
$r_{mean}$ values, to take in account this truncation, results in no
meaningful improvement in the accuracy of the photometric
redshifts. When comparing panels (a) and (d) there is a
remarkable difference between computing $r_{mean}$ within a fixed
luminosity range ($m^*-3 \le m_r \le m^*+3$, for instance) or using the
full survey limits (I considered all galaxies with $r \le 21$ for
panel d). The main effect is the overestimation of the counts at low
redshifts ($z < 0.15$), which leads to the flattening of the $r_{mean}
\times z$ relation in this regime. That happens because when using all
galaxies at $r \le 21$ for low-z clusters, we sample magnitudes that
are too faint in
comparison to the relevant regime of a cluster LF (such as $m^*-3 \le
m_r \le m^*+3$). Then, the $r_{mean}$ values become biased towards
higher values, with also a noticeable increase in the scatter. For
clusters at $z > 0.15$ there is no visible difference because $m^*+3$
is always close to $r = 21$ (or the survey limit is even extrapolated
for high-z systems). A similar discussion, but for richness
computation (instead of $r_{mean}$) is done in \citet{lop06}.
On what regards colors, the use of all
galaxies at $r \le 21$ has no large impact. We only see very few
clusters that have their median colors offset from the color tracks, 
increasing a little the scatter.

From the comparison of panels (b) and (e) we notice a clear trend for
underestimation of the {\it g-i} color at $z > 0.15$. The effect is
more pronounced at $z > 0.40$. The same effect is noticed for the 
{\it g-r} color (not shown in the plot). For {\it r-i} the effect is
not too drastic and we only see a mild underestimation at $z >
0.40$. Nonetheless, these results are very useful to show the relevance
of imposing a color cut (according to {\it u-r}) for measuring cluster
colors. By doing so, we can restrict the analysis to early type
galaxies, rendering the derived colors in good agreement to the
expectations of elliptical galaxies.

\subsection{Photometric redshifts of galaxy clusters}

I then proceed to derive empirical relations to estimate photometric
redshifts of clusters. That is done in a similar way to what is shown in
section 3.1 for LRGs, but here I use the mean magnitudes and median
colors of clusters, estimated as in the
left panels of figure 4. In other words, on top of the background
corrections I require galaxies to have $m^*-3 \le m_r \le m^*+3$ and also
impose a color cut (in {\it u-r}), as described above. For the 132
clusters of the training sample I derived the values of $r_{mean}$, 
($g-i$)$_{median}$ and ($r-i$)$_{median}$, which are shown in figure
4. An empirical relation between these three parameters and redshift
is then derived. Other colors are also obtained and will later be used
for comparison to the results based on the two above. This relation
can be expressed by 
\bea
\lefteqn{ z_{phot} =  {A + Br_{mean} + C(g-i)_{median} + D(r-i)_{median}}, } \, \nonumber \\
\label{eq:colourmagcuts}
\eea
analogous to the ones employed for LRGs. The derived coefficients are
\bea
\lefteqn{ A = -0.4424 \pm 0.0084} \, \nonumber\\
\lefteqn{ B = \,\,\,\, 0.0076 \pm 0.0006} \, \nonumber\\
\lefteqn{ C = \,\,\,\, 0.2382 \pm 0.0024} \, \nonumber\\
\lefteqn{ D = \,\,\,\, 0.2126 \pm 0.0042} \, \nonumber
\label{eq:coefficients}
\eea

To assess the accuracy of the photo-z estimator this relation is
applied to all 512 clusters with known
spectroscopic redshifts (section 4.1). However, as we would do with
clusters with unknown redshifts, we have to start the procedure with a guess
redshift and iterate it until the 
photometric redshift difference between two iterations is less than
0.01. A maximum of 10 iterations is allowed. Convergence is not found
for only 2 of the 512 clusters. This iterative procedure is necessary
as we sample 0.5 $h^{-1}$ Mpc and $m^*-3 \le m_r \le m^*+3$ for each
cluster, but we do not know what the redshift of the cluster is (used
to determine this radius and luminosity range for galaxy selection). So,
we start with a guess value ($z_{guess} = 0.15$), compute the mean
magnitude and median colors and apply the empirical relation obtained
above. Then we use the new redshift estimate to repeat the procedure
until convergence is achieved. The $z_{phot} \times z_{spec}$
comparison is shown on the top panel of Figure 5, while the weighted
residuals is exhibited in the lower panel. Clusters with
$|\delta_{z0}| < 0.10$ (see definition in $\S 3$) are shown as filled
circles (489 or 96\% of the 512). The remaining have open circles. If
I had used more than 132 clusters
(for instance, all the 512) to derive the photometric redshift
empirical relation the final results would have a slightly increased
scatter. That is due to the fact that some clusters
have spectroscopic redshifts derived from a small number of galaxies.
That is why it is important to use a clean sample to train the
calibration (\S 4.1).

\begin{table}
\begin{tabular}{lccccc} 
\hline
\multicolumn{6}{c}
{Photometric redshift errors of clusters from different relations} \\ 
\multicolumn{6}{c}
{} \\
Relation & $\mu$ & $\sigma$ & $\mu_0$ & $\sigma_0$ & Fraction(\%) \\ 
\hline
giri & -0.0021 & 0.026 & -0.0015 & 0.023 & 95.1 \\
rgrri & -0.0056 & 0.028 & -0.0052 & 0.025 & 95.3 \\
rgiri & -0.0046 & 0.026 & -0.0040 & 0.023 & 94.9 \\
grgiri & -0.0023 & 0.026 & -0.0021 & 0.023 & 95.1 \\
rgiri ($r \le 21$) & -0.0021 & 0.028 &  -0.0021 & 0.025 & 95.1 \\
rgiri (no {\it u-r} cut) & -0.0021 & 0.029 &  -0.0018 & 0.026 & 93.9 \\

\hline
\multicolumn{6}{c}
{Clusters with $z \le 0.30$} \\ 
\hline
giri & -0.0020 & 0.024 & -0.0019 & 0.022 & 95.3 \\
rgrri & -0.0069 & 0.025 & -0.0065 & 0.024 & 95.3 \\
rgiri & -0.0050 & 0.024 & -0.0048 & 0.022 & 94.9 \\
grgiri & -0.0023 & 0.025 & -0.0025 & 0.023 & 95.3 \\
rgiri ($r \le 21$) & -0.0025 & 0.026 &  -0.0029 & 0.024 & 95.4 \\
rgiri (no {\it u-r} cut) & -0.0024 & 0.026 &  -0.0024 & 0.024 & 96.1 \\

\hline
\multicolumn{6}{c}
{Clusters with $z > 0.30$} \\ 
\hline
giri & -0.0023 & 0.037 &  0.0016 & 0.030 & 92.9 \\
rgrri &  0.0051 & 0.041 &  0.0052 & 0.031 & 94.6 \\
rgiri & -0.0007 & 0.037 &  0.0026 & 0.030 & 94.6 \\
grgiri & -0.0022 & 0.036 &  0.0016 & 0.030 & 92.9 \\
rgiri ($r \le 21$) & 0.0013 & 0.042 &  0.0042 & 0.033 & 92.9 \\
rgiri (no {\it u-r} cut) & 0.0010 & 0.048 &  0.0045 & 0.041 & 76.8 \\

\hline
\end{tabular}
\caption{The mean, standard deviation and fraction of clusters with a valid
photometric redshift ($z_{phot} > 0$ and $|\delta_{z0}| < 0.10$). All rows 
show the results of polynomial fits obtained using different
parameters. In the first row I list the results when using the colors
($g-i$) and ($r-i$). Those based on the $r$ magnitude and colors
($g-r$) and ($r-i$) are show in the second row. In the third row the
results represent the use of the $r$ magnitude and colors ($g-i$) and
($r-i$). The fourth row has the results when using
the colors ($g-r$), ($g-i$) and ($r-i$). Fifth row exhibits the results
obtained with colors $r_{mean}$, (g-i)$_{median}$ and (r-i)$_{median}$,
but not being restricted to galaxies at $m^*-3 \le m_r \le m^*+3$. All
galaxies at $r \le 21$ are used in this case. In the sixth row the
results are again for $r$ magnitude and colors ($g-i$) and ($r-i$),
but the galaxies selected have no {\it u-r} cut applied. In the
continuation of the table, the same type of information is also given
for low ($z \le 0.30$) and high redshift clusters ($z > 0.30$).}
\label{tab:errors_clusters}
\end{table}

\begin{figure}
\includegraphics[height=0.50\textwidth,angle=0]{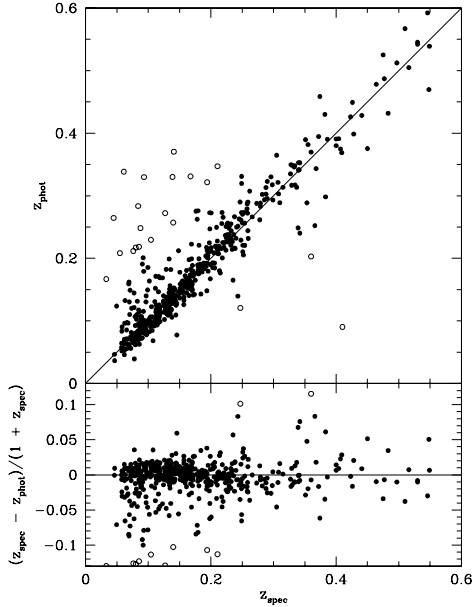}
\caption{Comparison between $z_{phot}$ and $z_{spec}$ using the empirical
relation based on the $r_{mean}$, ($g-i$)$_{median}$ and 
($r-i$)$_{median}$ colors 
(upper panel). In the lower panel the residual 
($\delta_{z0} = (z_{spec} - z_{phot}$)/($1+z_{spec}$)) is shown. The solid
lines indicate the Y = X result on top and the zero residual 
in the bottom panel.}
\label{fig:zspeczphot}
\end{figure}

In Table 3 I summarize the results obtained for relations based on
different combinations of mean $r$ magnitude and median
colors. Analogously to what was done for LRGs, I only use clusters with
$|\delta_z|$ or $|\delta_{z0}| < 0.10$ when computing $\mu$ and
$\sigma$ (or $\mu_0$ and $\sigma_0$). Had I included the gross
outliers, the standard deviation for the
$rgiri$ relation (third row of Table 3) is raised from 0.026 and
0.023 to 0.045 and 0.039 (values of $\sigma$ and $\sigma_0$,
respectively). Note that the fraction  of outliers is only $\sim
5\%$. See also the next section for some considerations
regarding accuracy.

The results in this table are first shown for
all clusters and then for those at $z \le 0.30$ and $z > 0.30$. On
each case, the first four rows list the results considering galaxies
at $m^*-3 \le m_r \le m^*+3$ and with the {\it u-r} color cut
applied. The fifth row exhibits the values when not imposing a fixed
luminosity range, while the last row has the results without a color
cut. When considering all clusters (but only the first four rows), we see
no large differences, except for the values achieved for the $rgrri$
relation ($r_{mean}$, (g-r)$_{median}$ and (r-i)$_{median}$) which
have $\sigma$ (or $\sigma_0$) a little higher than the rest. The same
is also true for low redshift clusters (which dominate the sample) and
the high redshift ones. Actually, the fraction of clusters at high-z
with a valid redshift decreases a little for the $giri$ and
$grgiri$ cases. Considering that, I decided to adopt the redshifts
obtained with $r_{mean}$, (g-i)$_{median}$ and (r-i)$_{median}$, as
they have fewer outliers and low dispersions at all redshifts. The
coefficients for equation 11 are listed above. These results show the
importance of using $r_{mean}$ and (g-i)$_{median}$ for photometric
redshift estimation at $z < 0.55$ in SDSS. The use of (g-r)$_{median}$
leads to an increased scatter (mainly at high-z), while using only
colors (without $r_{mean}$) gives lower completeness at high-z.

On what regards the
results without the luminosity or color constraints (rows 5 and 6),
they are noticeably worst than when these constraints are
enforced. When not applying the luminosity restriction the results are
at a similar level than what is obtained in the $rgrri$ case (except
for the number of outliers at high-z). Similar values are also derived
for the case where no color cut is applied, but only when considering
low-z clusters. At $z > 0.30$ the standard deviation and fraction of
outliers achieved without the color cut is the highest among all. That
tells us how important is the selection of early type galaxies for
cluster photo-z estimation at higher redshifts. Without this
pre-selection the cluster regions will likely be contaminated by
lower redshift sources or, more importantly, by blue galaxies (which
have an increased fraction at higher redshifts). Thus increasing the final
error and biasing the
results to low values. In other words, a simple background correction
is not enough, even considering that we are working with a small
aperture (0.50 $h^{-1}$ Mpc).

\subsection{Considerations about photometric redshift accuracy}

It is important to mention that the redshift accuracy achieved in this
work is lower than what other authors found with SDSS data. I
estimated $\sigma = 0.024$ for $z \le 0.30$, while \citet{got02} had
estimated uncertainties of $\sigma = 0.015$ and \citet{koe07} of
$\sigma = 0.01$ at $0.10 < z < 0.30$. Note that \citet{got02} show
residuals for clusters at $z > 0.08$. The inclusion of lower redshift
clusters in our sample ($z > 0.05$) helps increasing the scatter, but
by no means can explain the difference to other results.

\citet{koe07} estimate redshifts as part of the selection of
clusters. Each cluster has assigned to it the redshift of a galaxy
(brightest cluster galaxy, BCG)
that maximizes the likelihood of representing a cluster
center. \citet{got02} have
the redshift estimates done after cluster detection (with the {\it cut and
enhance} method). Their estimates are based on an early version of
the maxBCG technique, and are not identical to the ones from
\citet{koe07}. For a given redshift, they start counting the number of
galaxies within the cluster detection radius brighter than
$M_{r}$=-20.25 and within $\pm$1 mag in $g-r$ around the color
prediction for elliptical galaxies \citep{fuk95}. The procedure is
repeated for several redshifts in steps of $\delta z$= 0.01. After
the background is taken in account on each bin, the redshift of the
bin with the largest number of galaxies is considered the cluster
estimated redshift. 

In this work redshifts are obtained through the application of an
empirical relation to $r_{mean}$, (g-i)$_{median}$ and
(r-i)$_{median}$. The process is started with a guess redshift
($z_{guess} = 0.15$) and is iterated until convergence is
achieved. For the 512 clusters used here the photometric redshift
accuracy is simply given by the comparison of the measured
spectroscopic redshift and the photometric estimate. So, it is clear
that this process is not guided whatsoever. For
comparison, \citet{koe07} estimate the accuracy of their estimates by
selecting all clusters from their catalog that have spectroscopic
redshifts for their BCGs. For those, they compare $z_{phot}$ and
$z_{spec}$, where $z_{phot}$ is the cluster photo-z estimated from the
maxBCG algorithm and
$z_{spec}$ is the spectroscopic redshift measured for the BCG galaxy in
question. That seems a fair comparison, but the authors also
recognize that $\sim 16\%$ of their clusters suffer from projection
effects, which could affect cluster redshift estimates based on
several galaxies (and not only the BCG). On the other hand,
\citet{got02} seem to guide the
comparison of $z_{phot}$ and $z_{spec}$. In their words: ``the
redshift of the SDSS
spectroscopic galaxy within the detected radius and with nearest
spectroscopic redshift to the estimation is adopted as the real redshift''.
Such procedure obviously biases the comparison to small redshift
offsets, based on a single galaxy.

I decide to estimate the residuals in a similar way to what was
done by \citet{koe07} and \citet{got02}. For that purpose I did not
consider the 'Main' flux-limited sample of SDSS. Instead, I used only
the $197,956$ LRGs, with spectra available, selected in $\S 3$. In the
first case I selected
the nearest galaxy to the cluster center (within a maximum aperture of
60$''$). If the LRG selected is close to the center it might be the
BCG of the cluster. However, it is important to note that I do not
make any magnitude or color requirement for that selection. This
simple approach results in $\sigma = 0.021$ (or $\sigma_0 = 0.017$)
for the full redshift range of the clusters used here. When I restrict
the sample to clusters at $0.10 \le z \le 0.30$ I find $\sigma =
0.014$ (or $\sigma_0 = 0.011$). The last results are closer
to \citet{koe07} for the same redshift interval. For the full sample
there are 111 ($\sim 23\%$) clusters (out of 489 with a valid photo-z),
while there are 71 at $0.10 \le z \le 0.30$.

To perform a comparison to \citet{got02} I did something similar to
what they did. The only difference is to use a radius of 0.50 $h^{-1}$
Mpc, instead of the ``detection'' radius available within their
catalog. That should not result in meaningful differences as you do
not want to select a galaxy that is too far from the cluster
center. So, within 0.50 $h^{-1}$ Mpc I select the LRG with the closest
spectroscopic redshift to the value of $z_{phot}$ for each
cluster. For 301 clusters ($\sim 62\%$) there is at least one LRG
inside 0.50 $h^{-1}$ Mpc. Out of those, 206 are at $0.10 \le z \le
0.30$. I found $\sigma = 0.018$ (or $\sigma_0 = 0.017$) for the full
sample and $\sigma = 0.017$ (or $\sigma_0 = 0.015$) at $0.10 \le z \le
0.30$.

One problem with these tests is the fact that correlating the value of
$z_{phot}$ for a cluster with $z_{spec}$ for a single galaxy may lead
to wrong matches due to projection effects. As these clusters
represent a combination of objects from different catalogs in the
literature they are not supposed to have their centroid perfectly
matched with a BCG (substructure can affect the centroid
determination). That is not the case of the catalog from
\citet{koe07}. So, if their BCG has an spectroscopic observation and
their code works properly, they will have a good correlation between
$z_{phot}$ and $z_{spec}$ (except, perhaps, for clusters with strong
projection effects). Nonetheless, it is encouraging that the two tests
above result in an improved accuracy. 

To minimize the influence of projection effects I also estimated the
accuracy in a third way. For each cluster I select all LRGs within
360$''$ from the cluster center. From those, I check if the galaxy
closest to the centroid is at a maximum distance of 120$''$. If it is,
I assume the redshift of this LRG as a reference ($z_{ref}$). Then, from
all the other galaxies (LRGs) selected within 360$''$ I take those
that have a maximum redshift difference $|z_{lrg} - z_{ref}|$ of
  0.030. If I end up with at least three galaxies I take the mean of
  these redshifts to be the value of $z_{spec}$. When comparing these
  to $z_{phot}$ I find $\sigma = 0.016$ (or $\sigma_0 = 0.018$) for
  the full redshift range and $\sigma = 0.011$ (or $\sigma_0 = 0.010$)
  at $0.10 \le z \le 0.30$. There are 125 clusters ($\sim 26\%$) in
  the full sample and 81 in the restricted redshift interval. Note
  that this process is not guided whatsoever. I use an aperture that
  does not scale with redshift (360$''$) and check if there are at
  least three LRGs at the same redshift, taking as reference the
  redshift of the LRG closest to the cluster center (within
  120$''$). As I impose a minimum number of three galaxies, projection
  effects are minimized and we can see that the accuracy is greatly
  improved, reaching the 0.01 level found by \citet{koe07}.

\subsection{Application of the photometric redshift estimator}

As an application of the empirical photometric redshift relation
obtained in $\S 4.3$ I used SDSS data to derive new redshift estimates
for the supplemental version of the Northern Sky Optical Cluster
Survey (NoSOCS, \citealt{lop04}). This cluster catalog contains candidates to
$z \sim 0.50$, but the redshift estimates were based in a simple
magnitude-redshift relation. For that project the magnitude limit
adopted was $r = 21$. Thus, due to the large photometric errors, the
use of the $g-r$ color was not possible with data from the Digitized
Second Palomar Observatory Sky Survey (DPOSS). For details see
\citet{lop04}. 

This supplemental version of NoSOCS comprises 9956 cluster candidates over 
2,700 square degrees. I have recently searched the data release 5 of SDSS to
see which clusters already have SDSS data. After some catalog cleaning and
inspection of all cluster regions from SDSS I found that 7409 NoSOCS
clusters are well imaged. I then derived new photometric redshifts for
these systems using the relation obtained in this work. Actually, I
got new redshifts without re-centering and also considering new
coordinates, referred as luminosity weighted positions
\citep{lop06}. As expected,
the new redshift distribution has most of clusters at $z < 0.4$ with a
tail extending to $z \sim 0.55$. Details about this updated version of
NoSOCS will be given in future work, where substructure and
superposition effects will be investigated. For the lower redshift
clusters ($z < 0.1$) the updated version will also provide measures of
velocity dispersion, virial radius and mass; and this cluster subset will
be used for comparison to X-rays, as well to study scaling relations.

\subsection{Groups in redshift space with photometric estimates}

In the DR5 it is also possible to get photometric redshift estimates
of nearly all galaxies in SDSS. Two sources of redshifts are
provided. They are simply called \texttt{photoz} and
\texttt{photoz2}. The first one is based on template fitting using the
technique of \citet{csa03}. The second version is provided by the
UChicago/Fermilab/NYU group and is based on neural networks, with
implementation similar to \cite{col04}. For the results shown below I
used \texttt{photoz} \citep{csa03}.

For the 512 clusters from the training sample I selected photometric
redshifts for all galaxies from DR5. The same information
was also derived for
a region of $\sim 400$ square degrees. The boundaries of this area are
$168.4 < \alpha < 191.6$ and $20.0 < \delta < 40.0$, and were chosen to
avoid regions close to very nearby clusters, such as Virgo and
Coma. This sample will work as a large ``sky'' photometric redshift survey.

I investigated the possibility to identify groups in redshift space
simply using these galaxy photometric redshifts. I proceed as
follows. First, for each cluster, I start identifying groups in
redshift space. For that purpose, I use the gap-technique described in
\citet{kat96} and \citet{ols05}, which identifies gaps in the redshift
distribution that are larger than a given value to separate
groups. The gap size adopted is $\Delta z = 0.005(1+z)$ \citep{ols05},
which is approximately 1,500 $km/s$ in the restframe. I considered the
photometric redshifts available in the training sample to select all
galaxies within 0.50 $h^{-1}$ Mpc of the cluster center. When
performing galaxy selection, the
only requirement I make is that the photo-z of the galaxy should be greater
than 0.01 (to avoid very nearby structures and failures within the
SDSS estimates). After applying the gap-technique, a number of groups
in $z$-space is identified for each cluster

The next step is to assess the significance of each of these groups. For that I
consider the area of 400 square degrees described above. For each
group I draw 1000 sets of galaxies from the 400 square degrees
catalog. These sets have the same number of galaxies as in the cluster
region where the group was identified. The gap-technique is applied
exactly as before and then I check the probability of finding groups
with at least the same number of galaxies at the redshift of the
original group. A field group is considered if its redshift is within $\pm
0.005$ of the group identified in the cluster region. The significance
is given by the difference between one and the achieved
probability. I only consider groups that are significant at the 99\% level. 

From all the significant groups I select the one that has the smallest
redshift difference to the spectroscopic value of the cluster in
question. This group should also have at least three member galaxies
(most have many more) and be found within 3$'$ of the cluster
center. I have also ran this group
identification procedure with one slight modification, which is the gap
size. That was modified to $\Delta z = 500(1+exp(-(N-6)/33))/c$,
where N is the number of galaxies found in the redshift survey of a
cluster \citep{ada98}, and $c$ is the speed of light in $km/s$. 

In Figure 6 I summarize the results obtained for the two gap sizes
adopted. In the bottom
panel (b) the gap size considered is $\Delta z = 0.005(1+z)$,
being $\Delta z = 500(1+exp(-(N-6)/33))/c$ in the upper panel (a). The
total number of clusters identified
and the residuals between $z_{spec}$ and $z_{phot}$ are: (a) 298 and
$\sigma = 0.011$ (or $\sigma_0 = 0.009$); (b) 414 and $\sigma = 0.014$
(or $\sigma_0 = 0.013$). 

\begin{figure}
\includegraphics[height=0.50\textwidth,angle=0]{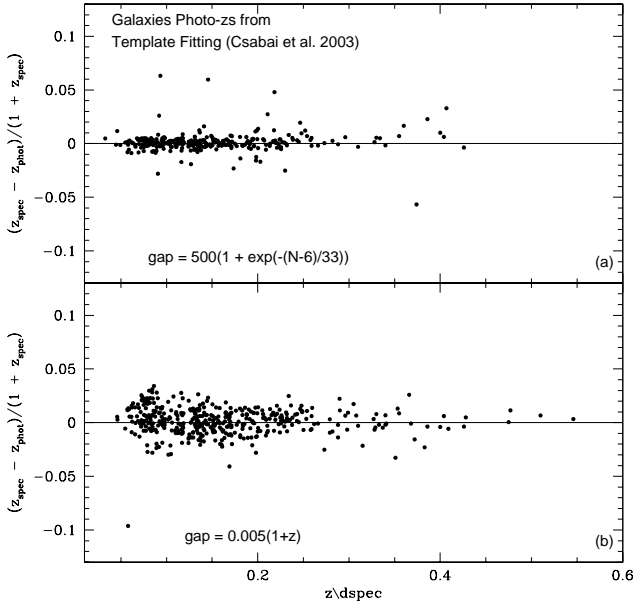}
\caption{Residual plots between spectroscopic redshifts of training
sample clusters and the closest group identified in redshift space
using photometric redshifts from SDSS. From the significant groups I
select the one with smallest redshift difference, at least three
member galaxies and within $3'$ of the cluster center. The top panel
consider the gap size $\Delta z = 500(1+exp(-(N-6)/33))/c$, while for
the bottom panel I use $\Delta z = 0.005(1+z)$. Details are given in
the text.}
\label{fig:zspeczphot}
\end{figure}

It is interesting to see that these galaxy photometric redshifts
provide values accurate enough to trace the local environment of
clusters. Similar results are achieved when I adopted the photometric
redshifts of LRGs ($\S 3$). From the inspection of figure 6 it is
clear that the identification of groups at higher redshifts ($z >
0.4$) becomes very hard. That is due to the poor sampling of the
luminosity function of clusters at this redshift interval for $r \le
21$. When using the gap size $\Delta z = 0.005(1+z)$ we identify more
clusters than with the other gap, but at the cost of having an
increased scatter.

Considering the larger number of blue galaxies at high redshifts the red
sequence in some clusters may be ill defined. So, it is important to consider
alternative approaches to cluster detection, such as the selection based 
on galaxy photometric redshifts.

\section{Conclusions}

In this work I described the construction of a large catalog of
luminous red galaxies with photometric redshifts at $z < 0.70$ within
SDSS\footnote{The LRG catalog can be retrieved
from the electronic edition of this journal or by request to the
author.}. This catalog is based on a empirical relation to derive the
photo-zs. Such relation uses three bands only for achieving results as
accurate as those obtained by other methods (sometimes based in more
bands; \citealt{pad05, col07}). The comparison of photometric and
spectroscopic redshifts show no large systematic effects in the
redshift range probed, which suggests that this sample is suitable
for large-scale structure studies.

I have also investigated what the main systematics are in the
estimation of photometric redshifts of galaxy clusters at $z <
0.55$. That represents an improvement respective to some deep cluster
catalogs derived from SDSS. When these catalogs have accurate
photometric redshift estimates, these are truncated at $z < 0.44$
\citep{got02}. When they go a little further ($z = 0.5$) the estimates
are not as accurate \citep{kim02}. I also showed the relevance of
using the $g-i$ color, and to a lesser extent the mean $r$ magnitude,
to improve the photometric redshift accuracy, especially at high-z ($z
> 0.4$). Besides that, I show that on top of a ``traditional'' background
correction it is very important to select galaxies from a fixed luminosity
range and perform a careful selection of early type galaxies. In this
work, this pre-selection of red galaxies is made through a variable cut
(in redshift) in the $u-r$ color. The main advantage of this type of
selection is to only require that clusters should exhibit a population
of early type galaxies towards their cores. There is no need for these
galaxies to exhibit a narrow red sequence. This type of requirement is
very important for the cases where the photometric errors are large or
the red-sequence is still being formed, which could be the case at
high redshifts.

The results obtained for clusters are independent of the way these are
selected. So, the methodology described in this work should be valid
for any type of clusters, selected by different techniques and
wavelengths. The only requirement is to have the proper filters for
separating early and late type galaxies and to track the 4000 {\AA}
break. For SDSS data this method works for clusters at $z < 0.55$.
Other considerations are made regarding accuracy. I show
that the results shown here are in good agreement to previous works.

The empirical relation derived for clusters is applied to 7409
clusters from the NoSOCS supplemental catalog \citep{lop04} which are
found within SDSS. For these clusters I was able to update the
photometric redshift estimates, deriving more accurate values than
before (when using only magnitudes within the DPOSS data). This
cluster catalog, with the new redshifts, will be updated in a future
work, where substructure and superposition effects will be
investigated. This catalog is also being used to derive velocity
dispersions and mass estimates for the lower redshift systems ($z <
0.1$) and to investigate scaling relations in clusters.

Finally, I tried to identify groups in redshift space using
photometric redshifts of galaxies available in SDSS. I found that for
$\sim 60\%$ of clusters (mostly at $z < 0.4$) it is possible to clearly
identify a group
using only photometric redshifts of galaxies. When comparing $z_{spec}$
of clusters to the photometric redshift of the nearest group, $\sigma
= 0.011$. If a different gap size is employed when searching for the
groups the rate of identified systems increases to $\sim 80\%$, but
the accuracy is a little worst ($\sigma = 0.014$). This procedure represents
an alternative approach for cluster detection, based on on galaxy photometric 
redshifts.

\section*{Acknowledgments}

PAAL was supported by the Funda\c c\~ao de Amparo \`a Pesquisa do
Estado de S\~ao Paulo (FAPESP, processes 03/04110-3 and 06/57027-4).
I also thanks the referee for a careful reading and comments,
which helped the presentation of this work.

This research has made use of the NASA/IPAC Extragalactic Database
(NED) which is operated by the Jet Propulsion Laboratory, California
Institute of Technology, under contract with the National Aeronautics
and Space Administration.

Funding for the SDSS and SDSS-II has been provided by the Alfred
P. Sloan Foundation, the Participating Institutions, the National
Science Foundation, the U.S. Department of Energy, the National
Aeronautics and Space Administration, the Japanese Monbukagakusho, the
Max Planck Society, and the Higher Education Funding Council for
England. The SDSS Web Site is http://www.sdss.org/.

The SDSS is managed by the Astrophysical Research Consortium for the
Participating Institutions. The Participating Institutions are the
American Museum of Natural History, Astrophysical Institute Potsdam,
University of Basel, University of Cambridge, Case Western Reserve
University, University of Chicago, Drexel University, Fermilab, the
Institute for Advanced Study, the Japan Participation Group, Johns
Hopkins University, the Joint Institute for Nuclear Astrophysics, the
Kavli Institute for Particle Astrophysics and Cosmology, the Korean
Scientist Group, the Chinese Academy of Sciences (LAMOST), Los Alamos
National Laboratory, the Max-Planck-Institute for Astronomy (MPIA),
the Max-Planck-Institute for Astrophysics (MPA), New Mexico State
University, Ohio State University, University of Pittsburgh,
University of Portsmouth, Princeton University, the United States
Naval Observatory, and the University of Washington.




\label{lastpage}

\end{document}